\documentclass[preprintnumbers,aps,showpacs,nofootinbib,superscriptaddress,floatfix]{revtex4} % for linux

\usepackage{graphicx}
\usepackage{amssymb}
\usepackage{amsmath}
\usepackage{bm}
\usepackage{slashed}
\usepackage{datetime}
\usepackage{epsfig}
\usepackage{soul} % use this (many fancier options)

\newcommand{\tmdpdf}{\ensuremath{\tilde{f}}}
\newcommand{\tmdpdfk}{\ensuremath{f}}
\newcommand{\trans}[1]{\ensuremath{{\bm #1}_{T}}}
\newcommand{\gammae}{\ensuremath{\gamma_{\rm E}}}
\newcommand{\subT}[1]{\ensuremath{#1_{T}}}

\begin{document}
\allowdisplaybreaks[2]

\preprint{JLAB-THY-13-1693}

\title{
Evolution of the helicity and transversity Transverse-Momentum-Dependent
parton distributions
}

\author{Alessandro Bacchetta}
\email{alessandro.bacchetta@unipv.it}
\affiliation{Dipartimento di Fisica, Universit\`a di Pavia, and INFN Sez. di
  Pavia, via Bassi 6, I-27100 Pavia}

\author{Alexei Prokudin}
\email{prokudin@jlab.org}
\affiliation{Jefferson Lab, 12000 Jefferson Avenue, Newport News, Virginia
  23606, USA}

%\date{\today}


\begin{abstract}
 We examine the QCD evolution of the helicity and transversity parton
 distribution functions when including also their dependence on transverse
 momentum.
Using an appropriate definition of these polarized
 transverse momentum distributions (TMDs), we describe 
 their dependence on the factorization scale and rapidity
 cutoff, which is essential for phenomenological applications.
\end{abstract}

\maketitle
%%%%%%%%%%%%%%%%%%%%%%%%%%%%%%%%%%%%%%%%%%%%%%%%%%%%%%%%%%%%%%%
\section{Introduction}

Our understanding of the partonic structure of hadrons relies on the
study of parton distribution functions (PDFs) and their extensions. In the last
years, particular attention has been devoted to transverse-momentum-dependent
parton distribution functions (TMDs). Standard collinear PDFs are defined
through collinear factorization theorems and obey the well-known DGLAP
evolution equations~\cite{Altarelli:1977zs,Dokshitzer:1977sg,Lipatov:1974qm}.
TMDs are defined through transverse-momentum-dependent
factorization and obey different evolution equations \cite{Collins:1984kg,Ji:2004wu,Collins:2011zzd}.  Here, for
the first time we analyze these TMD evolution equations for two important
distributions: the helicity and transversity TMDs. 

Factorization theorems are cornerstones of our understanding of hadron
structure.  They describe
experimentally measured cross-sections
in terms of perturbatively calculable hard parts and universal
structures related to nonperturbative parton dynamics, e.g., PDFs or TMDs. 
Factorization leads to well-defined evolution equations for the nonperturbative
functions, which allows
us to relate experimental measurements at different hard scales and perform
global analyses 
of PDFs, TMDs, and the corresponding fragmentation functions. 

The foundations of TMD factorization and evolution date back to
Refs.~\cite{Collins:1981uk,Collins:1984kg}. However, important details related
to gauge invariance have
been clarified only in the last decade (see, e.g.,
\cite{Ji:2002aa,Ji:2004wu,Ji:2004xq,Belitsky:2002sm,Collins:2002kn,Collins:2004nx}).
The first poof of TMD factorization was provided by Ji, Ma, and Yuan in Refs.~\cite{Ji:2004xq,Ji:2004wu} while a complete definition of TMDs and rigorous proof 
of factorization has been recently presented by
Collins in
Ref.~\cite{Collins:2011zzd} and applied in
Refs.~\cite{Aybat:2011zv,Aybat:2011ge,Aybat:2011ta,Anselmino:2012aa}.  
Another definition has been 
proposed in the context of Soft-Collinear Effective Theory by Echevarria,
Idilbi, and Scimemi in Refs.~\cite{GarciaEchevarria:2011md,Echevarria:2012pw} 
(see
also the closely-related work of Cherednikov and Stefanis
Refs.~\cite{Cherednikov:2007tw,Cherednikov:2009wk}).  
The relation between the approach of Collins and that of Echevarria,
Idilbi, and Scimemi
has been analyzed in Ref.~\cite{Collins:2012uy}, with the conclusion that they
are essentially equivalent.
In this work, we use the definition of Collins~\cite{Collins:2011zzd} and work in the
framework of his TMD factorization approach.

The nonperturbative objects introduced in factorization theorems 
typically depend on a renormalization scale $\mu$  in the case of collinear PDFs,
 and a so-called rapidity cutoff
$\zeta$ in the case of TMDs. Physical quantities do not depend on these
artificial scales, but only on the experimentally measurable hard scale of the
process (e.g., the photon invariant mass in a Drell--Yan process).

For the description of a spin-1/2 target, eight independent TMDs can be
introduced (at leading twist) \cite{Kotzinian:1994dv,Mulders:1995dh,Boer:1997nt,Bacchetta:2006tn}. At present, there exist explicit formulas for
the evolution of two of them: the unpolarized TMD, $f_1(x,\trans{k})$, 
and the Sivers
function, $f_{1T}^{\perp}(x,\trans{k})$. Here we consider the helicity
distribution, $g_{1}(x,\trans{k})$ 
and the transversity distribution $h_1(x,\trans{k})$. They are closely related
  to the collinear PDFs $g_{1}(x)$ and
$h_1(x)$, whose collinear evolution is well known
(see, e.g., Refs.~\cite{Barone:1997fh,Vogelsang:1997ak,Hayashigaki:1997dn}).

TMD evolution is related to transverse-momentum resummation and
so called Collins-Soper-Sterman (CSS) formalism~\cite{Collins:1984kg}. For the
polarized case of interest here, studies are presented in
Refs.~\cite{Koike:2006fn}. We will clarify in which sense our results
correspond to the ones presented in the resummation literature.
 
%%%%%%%%%%%%%%%%%%%%%%%%%%%%%%%%%%%%%%%%%%%%%%%%%%%%%%%%%%%%%%%%%%
\section{TMD Evolution}

We choose a particular process, namely Semi Inclusive Deep Inelastic
Scattering (SIDIS).
We denote with $P$
and $S$ the momentum and spin vector of the hadron target, and with $P_h$ the
momentum of the detected hadron.  With a single
exchanged photon of momentum $q$, independent kinematic variables are:
$Q=\sqrt{-q^2}$, $x=Q^2/2P\cdot q$, $z=P\cdot P_h/P\cdot q$, and the virtual
photon's transverse momentum ${\bm q}_{\rm T}$ (in a hadron frame
where the measured hadrons have zero transverse momentum).

Details about TMD factorization and definitions are given in
Refs.~\cite{Collins:2011zzd,Aybat:2011zv}. Here we summarize only the most important points.
A SIDIS structure function
in the form derived by Collins
\cite{Collins:2011zzd} reads: 
\begin{align}
\label{eq:parton2}
   F_{UU,T}(x,z,{\bm q}_T^2,Q^2) &= \sum_a \mathcal{H}_{UU,T}^{a}(Q;\mu) \, 
      \int d^2 {\bm k}_{ T} \, d^2 {\bm p}_{ T} \, 
      f_1^a\big(x,{\bm k}_{ T};\mu,\zeta_F\big) \, 
      D_{1}^{a}\big(z,z {\bm p}_{ T};\mu,\zeta_D\big) \,
      \delta^{(2)}\big({\bm k}_{ T} + {\bm q}_{ T} - {\bm p}_{ T}\big)
\nonumber\\&
 + Y_{UU,T}\big(Q,\trans{q}\big) + \mathcal{O}\big(\Lambda / Q\big).
\end{align} 
Here, $\mathcal{H}_{UU,T}$ is the hard scattering part, 
$f_1^a(x,{\bm k}_{T})$ is the TMD PDF
for an 
unpolarized quark of flavor $a$ in an unpolarised proton, and $D_1^{a}(z,z
{\bm p}_{T})$ is the 
unpolarized fragmentation function.
The formula is similar to the
parton-model expression (see, e.g.,~\cite{Bacchetta:2006tn}), except for the
dependence of the functions on $\mu$ and $\zeta$, the presence of higher-order
terms in the hard scattering, and the presence of the correction term $Y$,
which serves the purpose of correcting the expression of the structure
function at large $q_T\approx Q$, where the expression in terms of TMDs is not applicable. 
Analogous formulas can be derived for
other structure functions containing
other TMD PDFs and TMD FFs, see Refs.~\cite{Bacchetta:2006tn,Aybat:2011ge,Boer:2011xd}. 
$\Lambda$ denotes a generic hadronic scale, e.g., $\Lambda_{\rm QCD}$ or $M$. 
Note that we work in an approximation in which light cone momentum fractions
that enter in the definition of Eq.~\eqref{eq:parton2}  and usual Bjorken
variable $x$ are equal to each other. 

To correctly define the TMD PDFs and FFs in Eq.~\eqref{eq:parton2}, it is more
convenient to define them in 
in transverse coordinate space ($\trans{b}$-space) and then Fourier-transform the
final result. The structure functions, as that in Eq.~(\ref{eq:parton2}), can be
written as Fourier
transforms of $\trans{b}$-space expressions (see Sec.~2.2 of
Ref.~\cite{Boer:2011xd}).

The proper definition of TMD PDFs requires the introduction of the so-called
unsubtracted TMD PDFs together with further unsubtracted functions (sometimes
called ``soft factors''). Both of them contain rapidity divergences that are
eventually canceled in the final definition of the the TMD PDFs.
    
The
unsubtracted TMD PDFs are defined as (dropping the flavor index)
\begin{multline}
\label{eq:def1}
\tmdpdf_1^{\rm unsub}\big(x,\trans{b};\mu,y_P - y_B\big)= \\
 {\rm Tr} \int \frac{d \xi^{-}}{2 \pi} \,e^{-i x P^+ \xi^-}\, 
   \langle P,S | \bar{\psi}(\xi/2) W\big(\xi/2,\infty,n_B(y_B)\big)^\dagger 
  \frac{\gamma^+}{2} W\big(-\xi/2,\infty,n_B(y_B)\big) \psi(-\xi/2) | P ,S \rangle_{c}
\end{multline}
where $\xi^\mu = (0^+, \xi^-, {\bm b}_T)$, and we denote the functions with
a tilde to indicate that they are defined in 
transverse coordinate space.  The rapidity of the parent hadron is denoted by
$y_P$.  An additional  
parameter $y_B$ is needed to regulate the light-cone divergences that result
from using exactly light-like Wilson lines. This parameter is ultimately set
to $-\infty$ in the final definition and the rapidity divergence is canceled
by corresponding rapidity divergences in the soft function that we shall define below,
see Ref.~\cite{Collins:2011zzd}. 
The subscript $c$ indicates that only connected diagrams
are included.  The $W(a,b;n)$ functions represent Wilson lines
(gauge links) from $a$ to $b$ along the direction of the
four-vector $n$. This direction is determined by the choice of the process~\cite{Collins:2002kn}. 
It is essential for calculations of evolution for 
T-odd functions (Sivers and Boer-Mulders functions), but does not affect the
discussion of the present paper, which focuses on T-even distributions. 
Light cone variables are defined such as 
$a^{\pm} = (a^0 \pm a^3)/\sqrt{2}$ so that $a\cdot b = a^+b^- + a^-b^+ - {\bm a_T}\cdot {\bm b_T}$. 
To obtain the expressions for the  helicity and transversity TMD PDFs,
the
 Dirac structure $\gamma^+$ should be replaced by $\gamma^+\gamma_5$, and
 $\gamma^+\gamma^i\gamma_5$, respectively.
 
For the proper definition of TMD PDFs, we need also to introduce 
an unsubtracted soft function that corresponds to the expectation value of a
Wilson loop:
\begin{equation}
\label{eq:soft}
  \tilde{S}_{(0)}(\trans{b};y_A,y_B) = 
  \frac{1}{N_c} \langle 0 |W(\trans{b}/2,\infty;n_B)^{\dagger} \, 
                    W(\trans{b}/2,\infty;n_A) 
                    W(-\trans{b}/2,\infty;n_B)
                    W(-\trans{b}/2,\infty;n_A)^{\dagger} | 0 \rangle.
\end{equation}

In both \eqref{eq:def1} and \eqref{eq:soft}, also transverse
gauge links at infinity should be included~\cite{Ji:2002aa}. 
 However, when Feynman gauge is used
their effects cancel in the final TMD
PDF. Therefore we have not indicated the extra
gauge links explicitly.
The soft factor contains also self-interaction divergences that 
cancel in the final definition of TMDs.

The complete definition of the TMD PDF in $b_T$-space, given in Refs.~\cite{Collins:2011zzd}, is
\begin{equation}
\label{eq:TMDPDFdef}
\tmdpdf_1(x,\trans{b};\mu,\zeta_F) = \tmdpdf^{\rm unsub}_1 \big(x,\trans{b};\mu;y_P - (-\infty)\big) 
\sqrt{\frac{\tilde{S}_{(0)}(\trans{b};+\infty,y_s)}{\tilde{S}_{(0)}(\trans{b};+\infty,-\infty) \tilde{S}_{(0)}(\trans{b};y_s,-\infty)}} Z_F \, Z_2.
\end{equation}
Here, the ``$\infty$'' arguments for the rapidity variables in the unsubtracted PDF and the soft factors are meant in the sense of a limit. 
All field operators are unrenormalized, and $Z_F$ and $Z_2$ 
are the PDF and field strength renormalization factors respectively. The soft factors on the 
right-hand side of Eq.~(\ref{eq:TMDPDFdef}) contain rapidity arguments $y_s$.  It is an 
arbitrary parameter which can be thought of as separating the plus and minus directions.

To introduce polarization effects, 
we denote the Bloch 3-vector for a spin-1/2 particle moving in $+z$ direction as
\begin{equation}
 \boldsymbol{\rho} = (\boldsymbol{\rho}_{T},\lambda)\; ,
\end{equation}
 where $\lambda$ is the helicity and $\boldsymbol{\rho}_{T}$ is transverse spin, so that in massless limit
one has

\begin{equation}
 \frac{1}{2} \sum_s u(p)\bar u(p) = \frac{1}{2} \slashed{p} \bigg(1-\lambda \gamma_5 - \sum_{j=1,2}\gamma_5 \rho_{T}^j \gamma^j  \bigg) \; .
\label{eq:sum}
\end{equation}

If we define the following 4-vector
\begin{equation}
 \rho_T^\mu \equiv (0^+,0^-,\boldsymbol{\rho}_{T})\;,
\end{equation}
then we can rewrite Eq.~\eqref{eq:sum} as

\begin{equation}
 \frac{1}{2} \sum_s u(p)\bar u(p) = \frac{1}{2} \slashed{p} \left(1-\lambda \gamma_5 + \gamma_5\slashed\rho_{T} \right) \; ,
\label{eq:sum1}
\end{equation}
this equation will prove useful when calculating Feynman diagrams. We also choose an appropriate Sudakov decomposition for vectors introducing
two light cone vectors:
\begin{align}
p \equiv (p^+,0^-,\trans{0}) \; ,\\
n \equiv (0^+,1^-,\trans{0}) \; .
\end{align}

The Fourier transform for TMD PDFs in $D=4-2\epsilon$ reads
 \begin{align}
\tilde f(x,\trans{b}) \equiv \int {d^{2-2\epsilon} \trans{k}} e^{-i \trans{b}\cdot\trans{k}}f(x,\trans{k})\; .
%\tilde f(x,\trans{b}) \equiv \int \frac{d^{2-2\epsilon} \trans{k}}{(2\pi)^{2-2\epsilon}} e^{-i \trans{b}\cdot\trans{k}}f(x,\trans{k})\; ,
\label{eq:fourier}
\end{align}
The inverse Fourier transform reads
\begin{align}
f(x,\trans{k}) \equiv \int \frac{d^{2-2\epsilon} \trans{b}}{(2\pi)^{2-2\epsilon}} e^{i \trans{b}\cdot\trans{k}}\tilde f(x,\trans{b})\; .
%\tilde f(x,\trans{b}) \equiv \int \frac{d^{2-2\epsilon} \trans{k}}{(2\pi)^{2-2\epsilon}} e^{-i \trans{b}\cdot\trans{k}}f(x,\trans{k})\; ,
\label{eq:fourier_inverse}
\end{align}

\begin{figure*}[h]
\centering
  \begin{tabular}{c@{\hspace*{2cm}}c}
    \includegraphics[width=5cm]{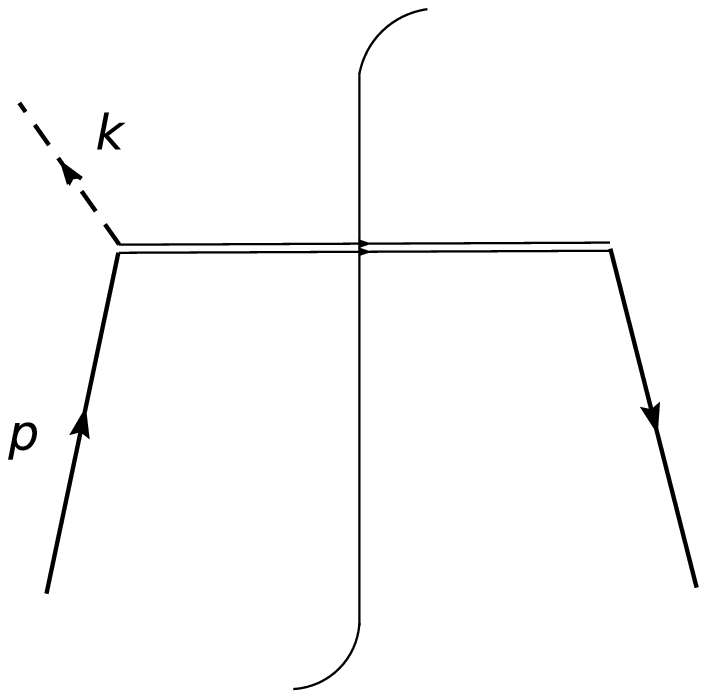}
    &
    \includegraphics[width=5cm]{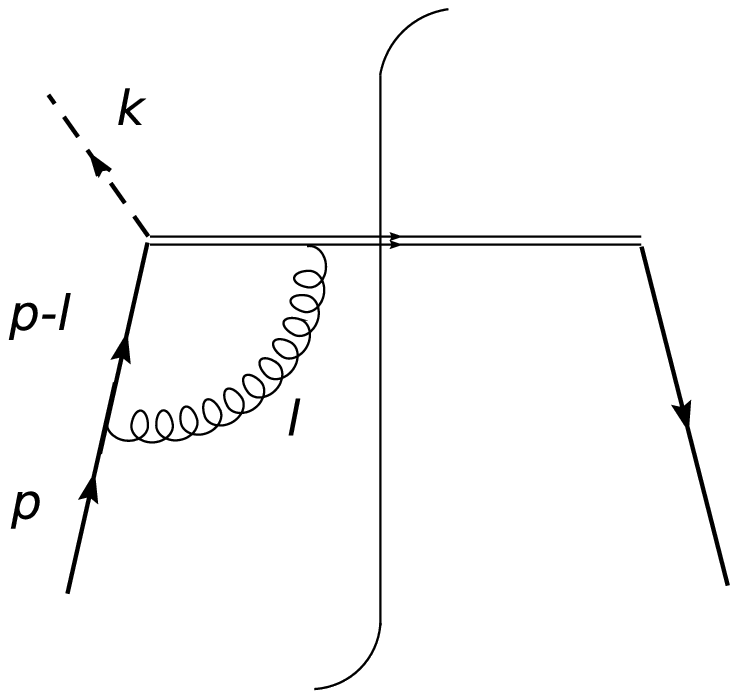}
  \\
  (a) & (b)
  \end{tabular}
\caption{ Tree level diagram (a) and example of virtual gluon emission diagram (b) for
  the calculation of TMD PDFs in quark target model, 
  $p$ is the momentum of the parent quark containing quark $k$.}
\label{fig:quarkmodel}
\end{figure*}

In order to address the problem of TMD evolution, we have first of all to
compute the TMD PDFs in a parton-target model (see
Fig.~\ref{fig:quarkmodel}.a). We will study distribution of a parton of type $j$
and momentum $k$ in a parton of type $i$ and momentum $p$. 
At tree level (Fig.~\ref{fig:quarkmodel}.a and the analogous case for gluons), the
results for the TMD PDFs we want to consider are
\begin{eqnarray}
% \tmdpdfk_{f/P}^{\rm unsub [0]}(x,\trans{k},S) &= &\delta(1-x) \delta^{(2)}(\trans{k})\; ,  \label{f}\\
% g_{f^\rightarrow/P^{\rightarrow}}^{\rm unsub [0]}(x,\trans{k},S) &=& \lambda \cdot \delta(1-x) \delta^{(2)}(\trans{k})\; ,  \label{g}\\
% h_{f^\upa rrow/P^{\uparrow}}^{\rm unsub [0]}(x,\trans{k},S) &=& \rho_{T,i} \cdot \delta(1-x) \delta^{(2)}(\trans{k})\; , \label{h}
% \tilde F_{f/P}^{\rm unsub [0]}(x,\trans{k},S) &= &\delta(1-x) \delta^{(2)}(\trans{k})\; ,  \label{f}\\
% \tilde G^{\rm unsub [0]}(x,\trans{k},S) &=& \lambda \cdot \delta(1-x) \delta^{(2)}(\trans{k})\; ,  \label{g}\\
% \tilde H^{\rm unsub [0]}(x,\trans{k},S) &=& \rho_{T,i} \cdot \delta(1-x) \delta^{(2)}(\trans{k})\; , \label{h}
 \tmdpdfk_{1 i}^{{\rm unsub} [0] j}(x,\trans{k}) &= &\delta(1-x)
 \delta^{(2)}(\trans{k})\delta^{j}_{i}\; ,  \label{f}\\
 g_{1 i}^{{\rm unsub} [0] j}(x,\trans{k},\lambda) &=& \delta(1-x) \delta^{(2)}(\trans{k})\delta^{j}_{i}\; ,  \label{g}\\
 h_{1 i}^{{\rm unsub} [0] j}(x,\trans{k},\boldsymbol{\rho}_T) &=& 
\delta(1-x) \delta^{(2)}(\trans{k})\delta^{j}_{i}\; , \label{h}
\end{eqnarray}
where $\tmdpdfk_{1}, g_1, h_1$ denote unpolarised distribution, helicity distribution and transversity distribution respectively.

%%%%%%%%%%%%%%%%%%%%%%%%%%%%%%%%%%%
%%%%%%%%%%%%%%%%%%%%%%%%%%%%%%%%%%%  

Including virtual gluon emission diagrams (see Fig.~\ref{fig:quarkmodel})
leads to multiplicative corrections to Eqs.~(\ref{f}, \ref{g}, \ref{h}) of the form
\begin{equation}
 \tmdpdfk_1^{\rm unsub [1]j}(x,\trans{k}) =  \tmdpdfk_1^{\rm unsub [0]j}(x,\trans{k}) \cdot \cal{C} \; ,
\end{equation}
where $\cal{C}$ is the appropriate result for the virtual gluon
loop:
\begin{align}
 {\cal{C}} = -i g^2 \mu^{2\epsilon} C_F \delta^j_{i}\int
\frac{d^{4-2\epsilon} l}{(2\pi)^{4-2\epsilon}}\frac{{\rm
    Tr}\left(\slashed{n}(\slashed{p}-\slashed{l})\slashed{n} \slashed{p}\right)}{ 4 (l^2+i\varepsilon)((p-l)^2+i\varepsilon)(n\cdot(p-k-l)+i\varepsilon)} \delta(p^+-k^+)\delta^{(2)}(\trans{k})\; .
\end{align}
  
The evolution of TMDs  follows from their definitions, 
Eqs.~(\ref{eq:TMDPDFdef}). 
The rapidity evolution (with respect to $\zeta_F$) is given by the  
Collins--Soper (CS) equation~\cite{Collins:1981uk}:
\begin{equation}
\label{eq:CSPDF}
\frac{\partial \ln \tilde{f}(x,\trans{b};\mu,\zeta_F)}{\partial \ln \sqrt{\zeta_F} } = \tilde{K}(\trans{b};\mu) 
\end{equation}
where the function $\tilde{K}(\trans{b};\mu)$ is defined as,
\begin{equation}
\label{eq:KPDF}
\tilde{K}(\trans{b};\mu) = 
\frac{1}{2} \frac{\partial}{\partial y_s} \ln \left( \frac{\tilde{S}(\trans{b};y_s,-\infty)}{\tilde{S}(\trans{b};+\infty,y_s)} \right).
\end{equation}
Note that the rapidity evolution depends only on the Soft factor, which is
independent of the polarization of the quark~\cite{Idilbi:2004vb}. 
Note that the equation contains 
$\tilde{S}(\trans{b})$ rather than $\tilde{S}_{(0)}(\trans{b})$.  Thus  
it is important to account for the UV renormalization factors $Z_F Z_2$ in
Eq.~\eqref{eq:TMDPDFdef}. 

The dependence on the scale $\mu$ arises from renormalization group equations 
for both $\tilde{f}(x,\trans{b};\mu,\zeta_F)$ and $\tilde{K}(\trans{b};\mu)$. 
They are
\begin{equation}
\label{eq:RGKPDF}
\frac{d \tilde{K}(\trans{b};\mu)}{d \ln \mu} = - \gamma_K\big(g(\mu)\big)
\end{equation}
and
\begin{equation}
\label{eq:RGPDF}
\frac{d \ln \tmdpdf(x,\trans{b};\mu,\zeta_F)}{d \ln \mu} = \gamma_F\big(g(\mu);\zeta_F /\mu^2\big).
\end{equation}
Here, $g(\mu)$ simply denotes the strong coupling with its dependence on the
scale. 
The functions $\gamma_K\big(g(\mu)\big)$ and $\gamma_F\big(g(\mu);\zeta_F /\mu^2\big)$ are the anomalous dimensions of 
$\tilde{K}(\trans{b};\mu)$ and $\tmdpdf(x,\trans{b};\mu,\zeta_F)$
respectively.
 The fact that $\gamma_K$ and $\gamma_F$  are
independent of $\trans{b}$ is due to the fact that UV divergence arises from
virtual gluon emission  diagrams only, see discussion in Ref.~\cite{Collins:2012ss}. 
As mentioned 
earlier, those diagrams give multiplicative factors, thus the overall result on evolution
does not depend on the polarization of the quark.
In conclusion, 
the results derived in Ref.~\cite{Collins:2011zzd,Aybat:2011zv} 
do not depend on the gamma matrix structure $\Gamma$ used to define the
specific polarized TMD, therefore they are universal for all polarization
states and allow us to
write down immediately the results for evolution of helicity and transversity
TMD PDFs.

There is a part where polarization is important. 
In the regime where $\trans{k}$ is large compared to the hadronic scale, but
still small compared to the hard scale (i.e., $\Lambda \ll
|\trans{k}| \ll Q$), TMDs can be calculated within a collinear 
factorization formalism~\cite{Collins:1982uw,Ji:2006ub,Bacchetta:2008xw}.
This means that 
when 
$\trans{b}$ is small but still larger than the inverse of the hard scale,
i.e., $1/Q \ll |\trans{b}| \ll 1/\Lambda$, 
Eq.~(\ref{eq:TMDPDFdef}) 
can be written as the
convultion of a perturbatively calculable 
hard scattering coefficient and an integrated PDF:
%\begin{multline}
\begin{equation}
\label{eq:smallb}
\tmdpdf_{1}^{j}(x,\trans{b};\mu,\zeta_F) = %\\
%=
\sum_{j'}\int_x^1\frac{d\hat{x}}{\hat{x}}  \tilde{C}_{j/j'}\big(x/\hat{x},\trans{b};\mu,\zeta_F\big) f_{1}^{j'}(\hat{x};\mu) %\\
+ \mathcal{O}(\Lambda b_T)\;,
\end{equation}
%\end{multline}
the sum  $j'$ goes over all quark and antiquark $q$, antiquark $\bar{q}$
flavors and gluon $g$. 
 The functions $f_{1}(\hat{x};\mu)$ are the ordinary integrated PDFs 
and the $\tilde{C}_{j/j'}(x/\hat{x},\trans{b};\mu,\zeta_F)$
are the hard coefficient functions. Similar expressions can be written for the helicity and
transversity distribution. The hard coefficients will be different and will be
denoted by $\Delta \tilde{C}$ for helicity and $\delta \tilde{C}$ for
transversity.  
We will explicitly calculate them for helicity and transversity distribution function.

Being independent of the type of initial hadron, 
the computation of the hard coefficients can be performed for the
parton-target case~\cite{Collins:2011zzd,Aybat:2011zv}. 
We can write perturbative results for the TMD distribution at small
$\trans{b}$ up the the first order
of perturbation expansion as (removing for convenience the dependence on the
scales)
\begin{equation}
\tmdpdf^{[1]j}_{1 i}(x,{\bm b}_T)=
\tilde{C}^{[1]}_{j/j'}(x/\hat x,{\bm b}_T) \otimes f^{[0]j'}_{1 i}(\hat x)
+ \tilde{C}^{[0]}_{j/j'}(x/\hat x,{\bm b}_T) \otimes f^{[1]j'}_{1 i}(\hat x)
\,. 
\end{equation}
The symbol $\otimes$ means the convolution from Eq.~\eqref{eq:smallb}.
The lowest order result for the hard coefficient is simply
\begin{equation}
\tilde{C}^{[0]}_{j/j'}(x/\hat x,{\bm b}_T)= \delta(1- x/\hat x) \delta_{j' j}.
\end{equation} 
Using the results for the lowest order of $f^{[0]}$ from
Eqs.(\ref{f}, \ref{g}, \ref{h}) integrated over $\trans{k}$ we obtain
\begin{equation}
\label{coeff}
\tilde{C}_{j/i}^{[1]}(x,{\bm b}_T)=\tilde{f}_{1 i}^{[1]j}(x,{\bm
  b}_T)-f_{1 i}^{[1]j}(x)\,. 
\end{equation}
This expression represents the recipe to compute the hard coefficients at order
$\alpha_S$. Analogous formulas hold for the hard coefficients $\Delta
\tilde{C}$ and $\delta \tilde{C}$ of the helicity and transversity
distributions. 

%  \begin{tabular}{c@{\hspace*{2cm}}c}
%    \includegraphics[width=5cm]{tree_level}
%%%%%%%%%%%%%%%%%%%%%%%%%%%%%%%%%%%
\begin{figure*}[t]
\centering
  \begin{tabular}{c@{\hspace*{2cm}}c}
 \includegraphics[width=5cm]{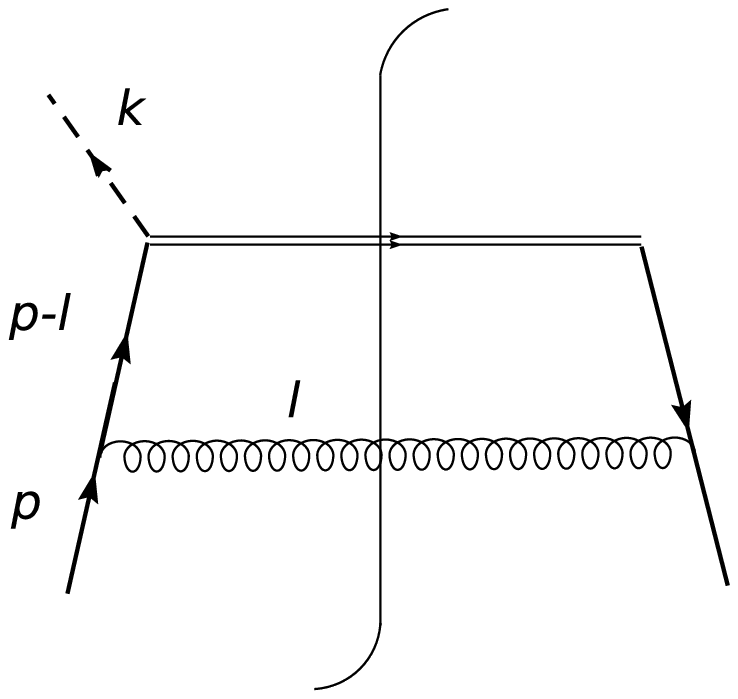}
    &
    \includegraphics[width=5cm]{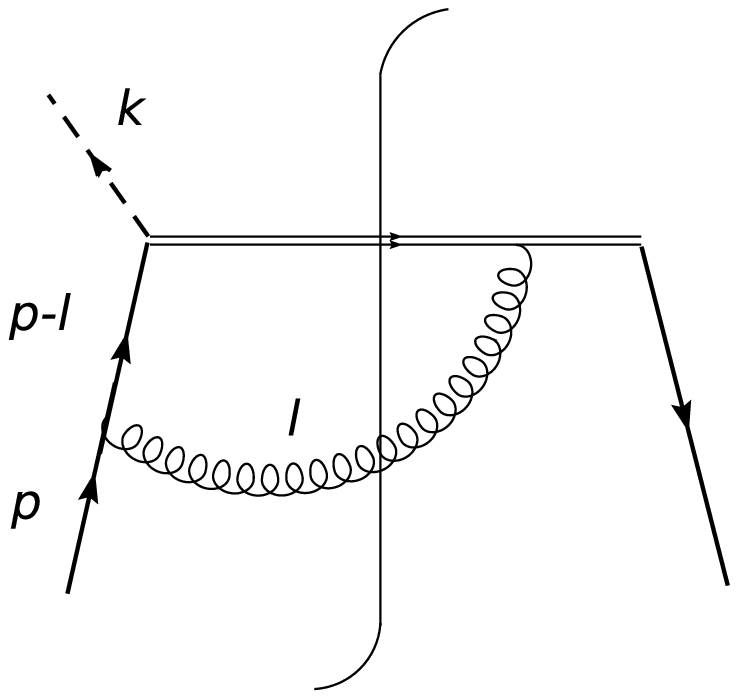}
  \\
  (a) & (b)
  \end{tabular}
\caption{ Real gluon emission diagrams (a) and (b) of TMD PDF in quark target
  model, $p$ is the momentum of the parent quark containing quark
  $k$. Hermitean conjugate diagrams have also to be taken into account.}
\label{fig:quarkmodelreal}
\end{figure*}
%%%%%%%%%%%%%%%%%%%%%%%%%%%%%%%%%%%  
 
For the TMD PDF of a quark in a quark we follow the steps of
Refs.~\cite{Collins:2011zzd,Aybat:2011zv}. To deal with eikonal propagators, 
we use the Feynman rules from
Refs.~\cite{Collins:1982uw,Collins:2011zzd}.
From diagrams (a) and
(b) in Fig.~\ref{fig:quarkmodelreal} we obtain
\begin{align}
\begin{split} 
f^{[1]j}_{1 i}(x,\trans{k})\Big\vert_a &= g^2 \mu^{2\epsilon} C_F \delta^j_{i}\int
\frac{d k^- d^{4-2\epsilon} l}{(2\pi)^{3-2\epsilon}}\frac{{\rm
    Tr}\left(\slashed{n}(\slashed{p}-\slashed{l})\gamma_\alpha\slashed{p}\left(1-\lambda
  \gamma_5 +\gamma_5 \slashed{\rho}_T\right)\gamma^\alpha
  )(\slashed{p}-\slashed{l})\right)}{ 4 ((p-l)^2+i\varepsilon)^2} 
\\ 
& \quad \times \delta^{(2)}(\trans{l}+\trans{k})\delta(p^+-l^+-k^+)\delta(l^2)\;
, \label{fa} 
\end{split} 
\\
\begin{split} 
f^{[1]j}_{1 i}(x,\trans{k})\Big\vert_b &= 
-g^2 \mu^{2\epsilon} C_F \delta^j_{i} \int \frac{d
  k^- d^{4-2\epsilon} l}{(2\pi)^{3-2\epsilon}}\frac{{\rm
    Tr}\left(\slashed{n}(\slashed{p}-\slashed{l})\slashed{n}\slashed{p}\left(1-\lambda
  \gamma_5 +\gamma_5 \slashed{\rho}_T\right)\right)}{ 4
  ((p-l)^2+i\varepsilon)(n\cdot l - i\varepsilon)} 
\\ 
& \quad \times \delta^{(2)}(\trans{l}+\trans{k})\delta(p^+-l^+-k^+)\delta(l^2)\; . \label{fb}
\end{split} 
\end{align}
The Dirac structure $\slashed{n}$ has to be replaced by $\slashed{n}\gamma_5$
and  $\slashed{n}\gamma^i\gamma_5$ for the helicity and
transversity distributions, respectively.

Calculations of integrals in Eqs.~(\ref{fa},\ref{fb}) go along the lines of Ref.~\cite{Collins:2011zzd}. After estimating the trace in ${4-2\epsilon}$ dimensions we evaluate the integral over $d l^-$ first by
closing the contour at infinity and using Gauchy's integral theorem. Then we evaluate the reminding $d l^+$ and $d^{2-2\epsilon}\trans{l}$ integrals
utilizing delta functions. Finally we compute $\overline{MS}$ counterterms by prescription from Ref.~\cite{Collins:2011zzd}.

Using  Eq.~\eqref{coeff} and results of Eqs.~(\ref{fa},\ref{fb})  along 
with the soft subtraction factors (see Appendix A of Ref.~\cite{Aybat:2011zv})
   for the TMD PDF for finding a 
quark  of flavor $j^\prime$ in a quark of flavor $j$ we find to order $\alpha_s$,

\begin{multline}
\tilde{C}_{j^\prime / j}(x,\trans{b};\mu;\zeta_F / \mu^2) = 
\delta_{j^\prime j} \delta(1 - x) 
+  \delta_{j^\prime j} \frac{\alpha_s C_{\rm F}}{\pi} 
\Bigg\{ \ln \left( \frac{2 e^{-\gammae}}{\mu \subT{b}} \right) 
\bigg( \frac{1+x^2}{1 - x} \bigg)_{+} + \frac{1}{2}(1 -x) + \\ 
+ \delta(1 - x) \left[  -  \ln^2 \left( \frac{2 e^{-\gammae}}{\mu
        \subT{b}} \right)  
+  \ln \left( \frac{2 e^{-\gammae} }{\mu \subT{b}} \right) 
\ln \left( \frac{\zeta_F}{\mu^2} \right) \right] \Bigg\} 
+ \mathcal{O}(\alpha_s^2)\; ,
\label{Cf}
\end{multline}

\begin{multline}
\Delta \tilde{C}_{j^\prime / j}(x,\trans{b};\mu;\zeta_F / \mu^2) = 
\delta_{j^\prime j} \delta(1 - x) 
+  \delta_{j^\prime j} \frac{\alpha_s C_{\rm F}}{\pi} 
\Bigg\{ \ln \left( \frac{2 e^{-\gammae}}{\mu \subT{b}} \right) 
\bigg( \frac{1+x^2}{1 - x} \bigg)_{+} + \frac{1}{2}(1 -x) + \\ 
+ \delta(1 - x) \left[  -  \ln^2 \left( \frac{2 e^{-\gammae}}{\mu
        \subT{b}} \right)  
+  \ln \left( \frac{2 e^{-\gammae} }{\mu \subT{b}} \right) 
\ln \left( \frac{\zeta_F}{\mu^2} \right) \right] \Bigg\} 
+ \mathcal{O}(\alpha_s^2)\; ,
\label{Cg}
\end{multline}
 
\begin{multline}
\delta \tilde{C}_{j^\prime / j}(x,\trans{b};\mu;\zeta_F / \mu^2) = 
\delta_{j^\prime j} \delta(1 - x) 
+  \delta_{j^\prime j} \frac{\alpha_s C_{\rm F}}{\pi} 
\Bigg\{ \ln \left( \frac{2 e^{-\gammae}}{\mu \subT{b}} \right) 
\bigg( \frac{2x}{1 - x} \bigg)_{+}  + \\ 
+ \delta(1 - x) \left[  -  \ln^2 \left( \frac{2 e^{-\gammae}}{\mu
        \subT{b}} \right)  
+  \ln \left( \frac{2 e^{-\gammae} }{\mu \subT{b}} \right) 
\ln \left( \frac{\zeta_F}{\mu^2} \right) \right] \Bigg\} 
+ \mathcal{O}(\alpha_s^2)\; .
\label{Ch}
\end{multline}
for unpolarised, helicity, and transversity TMDs respectively. The strong
coupling $\alpha_s$ is evaluated at a
scale $\mu$, and the number of active flavors is  e $N_f$. The usual $SU(N_c)$ color
factors are  $C_{\rm F}=(N_c^2-1)/(2N_c)$, $T_{f}=1/2$. 

%%%%%%%%%%%%%%%%%%%%%%%%%%%%%%%%%%%
\begin{figure*}[h]
\centering
     \includegraphics[scale=0.6]{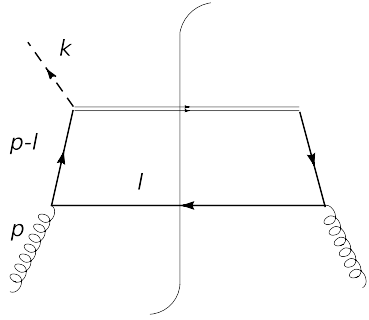}
\caption{ Quark in a gluon diagram for TMD PDF, $p$ is the momentum of the parent gluon containing quark $k$.}
\label{fig:quarkingluon}
\end{figure*}
%%%%%%%%%%%%%%%%%%%%%%%%%%%%%%%%%%%  

The calculation for quark in gluon follow the lines of
Refs.~\cite{Collins:2011zzd,Aybat:2011zv}. We utilize the diagram from
Fig.~\ref{fig:quarkingluon}. Note that at the order $\alpha_s$ there is no
contribution from soft factor subtraction and all the diagrams with gluon
attached to the Wilson line give zero 
if an appropriate choice of gluon polarization is made, see
Refs.~\cite{Collins:2011zzd}. 
Furthermore, the diagram from Fig.~\ref{fig:quarkingluon} gives zero for
transversity, since using the transversity projector $\gamma^+\gamma^i\gamma_5$ the Dirac trace contains odd number of Dirac matrices. This result is well
known in the literature.
 
Using Feynman rules from Ref.~\cite{Collins:2011zzd} we obtain for diagrams  in
Fig.~\ref{fig:quarkingluon}  
\begin{align}
f^{[1] j}_{1 g}(x,\trans{k}) &= - \frac{T_f g^2 \mu^{2\epsilon}} {4(1-\epsilon)} g_\perp^{\mu\nu} \int \frac{d k^- d^{4-2\epsilon} l}{(2\pi)^{4-2\epsilon}}
\frac{{\rm Tr}\left(\slashed{n}
  \slashed{k}\gamma_\mu(\slashed{k}-\slashed{p})\gamma_\nu
  \slashed{k}\right)}{ (k^2+i\varepsilon)^2}\delta\left((p-k)^2\right)
, \label{fgq} 
\\
g^{[1] j}_{1 g}(x,\trans{k})
&= - \frac{T_f g^2 \mu^{2\epsilon}} {4(1-3\epsilon)} \epsilon_\perp^{\mu\nu} \int \frac{d k^- d^{4-2\epsilon} l}{(2\pi)^{4-2\epsilon}}
\frac{{\rm Tr}\left(\slashed{n}\gamma_5 \slashed{k}\gamma_\mu(\slashed{k}-\slashed{p})\gamma_\nu \slashed{k}\right)}{ (k^2+i\varepsilon)^2}\delta\left((p-k)^2\right)  , \label{fgq2}
\end{align}
for unpolarised and helicity distributions accordingly. Here
\begin{align}
g_\perp^{\mu\nu} &\equiv -g^{\mu\nu} + \frac{1}{p^+}\left( p^\mu n^\nu + n^\mu p^\nu \right) \; ,\\
\epsilon_\perp^{\mu\nu} &\equiv  \frac{1}{p^+} \epsilon^{\alpha\beta\mu\nu} p_{\alpha} n_{\beta} \; .
\end{align}
 
In conclusion, for the gluon contributions we obtain
\begin{align}
\tilde{C}_{g / j}(x,\trans{b};\mu,\zeta_F / \mu^2) &=  \frac{\alpha_s T_{f}}{\pi} 
\left\{ \ln \left( \frac{2 e^{-\gammae}}{\mu \subT{b}} \right) 
\big( x^2 + (1-x)^2 \big)   
%\right. + 
%\nonumber \\ 
%\left. \vphantom{\left[ \ln \left( \frac{2}{\mu \subT{b}} \right) - \gammae \right] } 
+ x(1 - x)  \right\} + \mathcal{O}(\alpha_s^2).
\label{C1f}
\\
\Delta \tilde{C}_{g / j}(x,\trans{b};\mu,\zeta_F / \mu^2) &=  \frac{\alpha_s T_{f}}{\pi} 
\left\{ \ln \left( \frac{2 e^{-\gammae}}{\mu \subT{b}} \right) 
\left( 2x -1  \right) 
%\right. + 
%\nonumber \\ 
%\left.  \vphantom{\left[ \ln \left( \frac{2}{\mu \subT{b}} \right) - \gammae \right] } 
+ (1 - x)  \right\} + \mathcal{O}(\alpha_s^2).
\label{C1g}
\\
\delta \tilde{C}_{g / j}(x,\trans{b};\mu,\zeta_F / \mu^2) &= 0  + \mathcal{O}(\alpha_s^2).
\label{C1h}
\end{align}
for unpolarised, helicity, and transversity TMDs, respectively. Note that there are no contributions from Soft factor at
this order to quark in a gluon coefficient functions.

Eqs.~(\ref{Cg}, \ref{Ch})
and Eqs.~(\ref{C1g}, \ref{C1h}) represent the original results of this
paper. They 
allow us to write epxressions for the helicity and transversity TMDs that
fulfill TMD evolution equations, Eqs.~(\ref{eq:CSPDF}, \ref{eq:RGKPDF},
\ref{eq:RGPDF}), 
and have a behavior at high
transverse momentum that matches perturbative calculations.
The 
solution 
for 
a TMD for flavor $i$ can be written in a compact way as (see Refs.~\cite{Collins:2011zzd,Aybat:2011zv}) 
\begin{equation}   
\widetilde{f}_1^f(x,b_T;\mu,\zeta_F) =
\sum_i\bigl( 
\tilde{C}_{f/i}\otimes
f_1^i\bigr)(x,b_{\ast};\mu_b) 
e^{\tilde S(b_{\ast};\mu_b,\mu,\zeta_F)}
e^{g_K(b_T)\ln\frac{\sqrt{\zeta_F}}{\sqrt{\zeta_{f0}}}}
\hat{f}_{ {\rm NP}}^q(x,b_T) 
\label{e:TMDevol1}
\end{equation}
Analogous formulas hold for the
 helicity, ${g_1}$, or transversity, ${h_1}$, distributions. The sum goes over all
quark and antiquark flavors and include also gluon, $j=q,\bar q,
g$. Appropriate coefficient functions should be used in each case from
Eqs.~(\ref{Cf}, \ref{Cg}, \ref{Ch}),
$\otimes$ denotes the
convolution in longitudinal momentum fractions of Eq.~\eqref{eq:smallb}. 
The scale $\mu_b$ is chosen appropriately to ensure the optimal convergence of perturbative series.
The function $\hat{f}_{ {\rm NP}}$ 
denotes the nonperturbative part of the
TMD and has to be fitted to experimental data. In the literature, it is
usually parametrized as a Gaussian, although there is no fundamental reason
for this choice.

In order to be able to use Eq.~\eqref{eq:smallb} also at large $b_T$,
the so-called $b_{\ast}$ prescription can be introduced.
The function $b_{\ast}$ serves the purpose of freezing the value of $b_T$,
preventing it from becoming larger than a certain value and
avoid regions where the perturbative running coupling $\alpha_s$ becomes
divergent (at very large values of $b_T$, or equivalently, at very small
transverse momentum). A common choice is to set
\begin{equation} 
b_{\ast}\equiv \frac{b_T}{\sqrt{1+b_T^2/b_{\rm max}^2}},
\end{equation} 
but other functional forms can be explored, as well as different prescriptions
(e.g., the complex-$b$ prescription of
Ref.~\cite{Laenen:2000de,Kulesza:2003wn}). Any change in the prescription
should also be combined with a change of the nonperturbative function $\hat{f}_{ {\rm NP}}$.

The perturbatively calculable function $\tilde S(b_{\ast};\mu_b,\mu,\zeta_F)$ reads
\begin{align}
\tilde S(b_{\ast};\mu_b,\mu,\zeta_F) = \ln \frac{\sqrt{\zeta_F}}{\mu_b} \tilde K (b_{\ast};\mu_b) + \int_{\mu_b}^\mu \frac{d\mu'}{\mu'}
\left[ \gamma_F(g(\mu');1) - \ln \frac{\sqrt{\zeta_F}}{\mu'} \gamma_K (g(\mu')) \right] \; ,
\end{align}
and expressions for $\tilde K $,$\gamma_F$, and $\gamma_K$ at order $\alpha_S$ can be found in Appendix B of Ref.~\cite{Aybat:2011zv}.

We adopt also these other choices \footnote{Note that parameter $b_0$ is dimensionless, so
  that $b_0/b_{\ast}$ has the dimensions of energy, GeV.}
\begin{align}
\mu&=Q, 
&
\mu_b&=2 e^{-\gamma_E}/b_{\ast}\equiv b_0/b_{\ast},
&
\zeta_F &=Q^2,
&
\zeta_{F 0} &= Q_0^2.
\label{choice}
\end{align} 
Different options can be explored in order to test the sensitivity of the
final results to the scale choice. Note that $\tilde K (b_{\ast};\mu_b) = 0$ at this order with this choice.

Using Eq.~\eqref{choice} we find  to order $\alpha_s$ for Eqs.(\ref{Cf},\ref{Cg},\ref{Ch}),
\begin{align}
\tilde{C}_{j^\prime / j}(x,b_\ast;\mu_b ) & = \delta_{j^\prime j} \delta(1 -
x) +  \delta_{j^\prime j} \frac{\alpha_s C_{\rm F}}{2 \pi}  
\;(  1 -x ) + \mathcal{O}(\alpha_s^2)\, , \label{e:Cffinal}
\\
 \Delta  \tilde{C}_{j^\prime / j}(x,b_\ast;\mu_b )& =  \delta_{j^\prime j}
 \delta(1 - x) +  \delta_{j^\prime j} \frac{\alpha_s C_{\rm F}}{2 \pi}  
\;(  1 -x ) + \mathcal{O}(\alpha_s^2)\, , \label{e:Cgfinal}
\\
 \delta  \tilde{C}_{j^\prime / j}(x,b_\ast;\mu_b )& =  \delta_{j^\prime j}
 \delta(1 - x) +   \mathcal{O}(\alpha_s^2). 
\label{e:Chfinal}
\end{align}
and for Eqs.(\ref{C1f},\ref{C1g},\ref{C1h}),
\begin{align}
\tilde{C}_{g / j}(x,b_\ast;\mu_b ) & =    \frac{\alpha_s T_{f}}{\pi} 
\;x (  1 -x ) + \mathcal{O}(\alpha_s^2)\, , \label{e:Cgffinal}\\
 \Delta  \tilde{C}_{g / j}(x,b_\ast;\mu_b )& =   \frac{\alpha_s T_{f}}{\pi} 
\; (  1 -x ) + \mathcal{O}(\alpha_s^2)\, , \label{e:Cggfinal}\\
 \delta  \tilde{C}_{g / j}(x,b_\ast;\mu_b )& =    \mathcal{O}(\alpha_s^2).\label{e:Cghfinal}
\end{align}
for unpolarised, helicity, and transversity TMDs, respectively. 
One can see that
$\tilde{C}_{j^\prime / j} = \Delta \tilde{C}_{j^\prime / j}$ and the
difference between $\tilde{C}_{j^\prime / j}$ and $\delta\tilde{C}_{j^\prime /
  j}$  
is the absence of $\alpha_s$ contribution. 

At this point, 
a discussion on the comparision with the CSS literature is in order.
The restructuring of the
formalism in terms of TMD definitions makes it non-trivial to map them onto
the classic CSS components. However, 
we can check that the final result for the
structure functions match. For unpolarized DIS, results in the CSS formalism
are presented, e.g., in Ref.~\cite{Nadolsky:1999kb}. Polarized DIS has been
discussed in Ref.~\cite{Koike:2006fn}.
For instance, we can compare the results for the
unpolarized structure function of Eq.~(\ref{eq:parton2}). We need the
expression of the hard scattering, which 
has been reported in Ref.~\cite{Aybat:2011vb}
\begin{align} 
\mathcal{H}_{UU,T}^{a}(Q;\mu) 
&= 
e_a^2 \left(1 + \frac{C_F \alpha_s}{\pi} 
\left[\frac{3}{2} \ln \left( \frac{Q^2}{\mu^2} \right) - \frac{1}{2} \ln^2
  \left( \frac{Q^2}{\mu^2} \right) - 4 \right] \right).
\label{eq:hard}
\end{align} 

If we insert the expression for the unpolarized TMD, Eq.~\eqref{e:TMDevol1},
the analogous expression for the fragmentation function, Eq.~(31) of
Ref.~\cite{Aybat:2011zv}, 
and the hard scattering, Eq.~(\ref{eq:hard}), in the formula of the structure
fuction, 
Eq.~(\ref{eq:parton2}) and we adopt the choices of Eq.~(\ref{choice}), we
recover the standard CSS results for DIS at order $\alpha_S$ 
(see, e.g., Eq.~(45) of Ref.~\cite{Nadolsky:1999kb}, or
Eq.~(32) of Ref.~\cite{Koike:2006fn}). Other structure functions for polarized
DIS require the study of different TMD PDFs and FFs, together with a careful
assesment of the possibility to match the low and high transverse momentum
results (see, e.g., Ref.~\cite{Koike:2007dg,Bacchetta:2008xw,Kang:2011mr}). 
Since here we are concerned
only with helicity and transversity TMDs, results can be compared with those
of Ref.~\cite{Koike:2006fn}. 

Our results can be compared also to the Drell--Yan case. The unpolarized case
has been 
discussed in many  papers,
 e.g., 
Refs.~\cite{Collins:1984kg,Arnold:1990yk,Ellis:1997sc,Nadolsky:2000ky,
Bozzi:2010xn}). 
In
this case, however, the hard scattering of Eq.~(\ref{eq:hard}) 
has an extra $\pi^2/2$ in the
square brackets, as in Eq.~(6) of Ref.~\cite{Aybat:2011vb}). Our results can
be compared to those for doubly 
longitudinally
and transversely polarized Drell--Yann~\cite{Weber:1991wd,Kawamura:2005kj}. 

Apart from the full structure function, we can focus our
attention on the coefficient functions: our results expressed in
Eqs.~(\ref{e:Cffinal}, \ref{e:Cgfinal}, \ref{e:Chfinal}) 
and in Eqs.~(\ref{e:Cgffinal}, \ref{e:Cggfinal}, \ref{e:Cghfinal}) 
correspond to 
Eq.~(37, 38, 41, 42, 43) of Ref.~\cite{Koike:2006fn}. Differences occur only
for the terms with $\delta(1-x)$ and can be ascribed to the contribution from
the hard scattering. These contributions 
turn out to be the same for all three cases, which indicates that
the hard scattering is the same for the relevant structure functions. 
This has been already observed, albeit with a
different definition of the hard scattering, in
Refs.~\cite{Ji:2004xq,Kang:2011mr}. 
Since the hard scattering is different in different processes, 
in the CSS formalism 
the coefficients are different in DIS and Drell--Yan
scattering. This can be checked by comparing, e.g., Eq.~(38, 42) of
Ref.~\cite{Koike:2006fn} with the coefficient $\Delta c_q$ in Eq.~(15) of 
Ref.~\cite{Weber:1991wd}
and
Eq.~(38, 43) of
Ref.~\cite{Koike:2006fn} with the $\alpha_s$ part of 
Eq.~(8) of Ref.~\cite{Kawamura:2005kj}. The
difference between these results is due to the factor $\pi^2/2$ that we
discussed previously.

%%%%%%%%%%%%%%%%%%%%%%%%%%%%%%%%%%%%%%%
\section{Phenomenology}

%%%%%%%% FIG 4 %%%%%%%%%%%%%
\begin{figure*}[htb]
\psfig{file=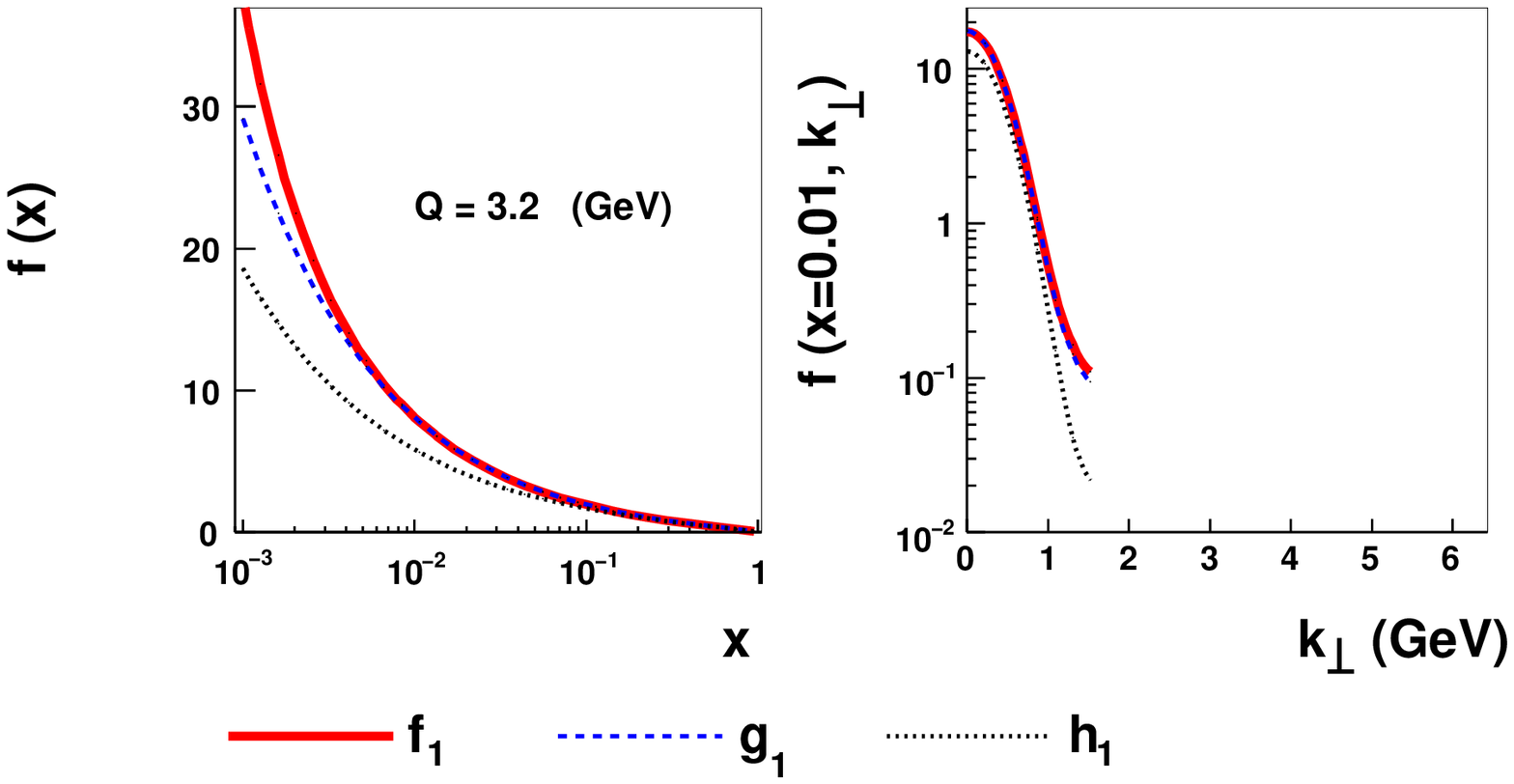, width=0.62\textwidth}
\vskip -2cm
\psfig{file=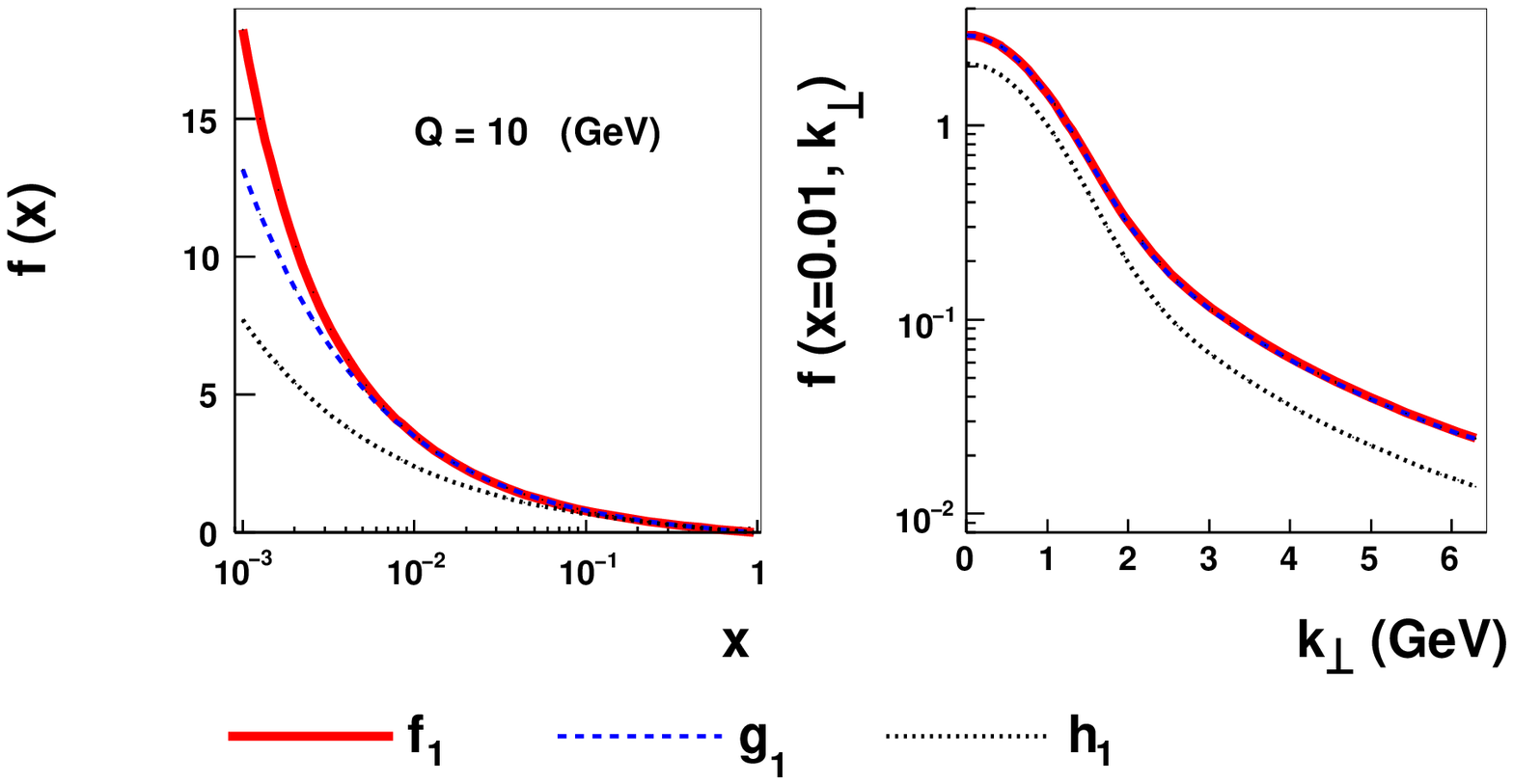, width=0.62\textwidth}
\vskip -2cm
\psfig{file=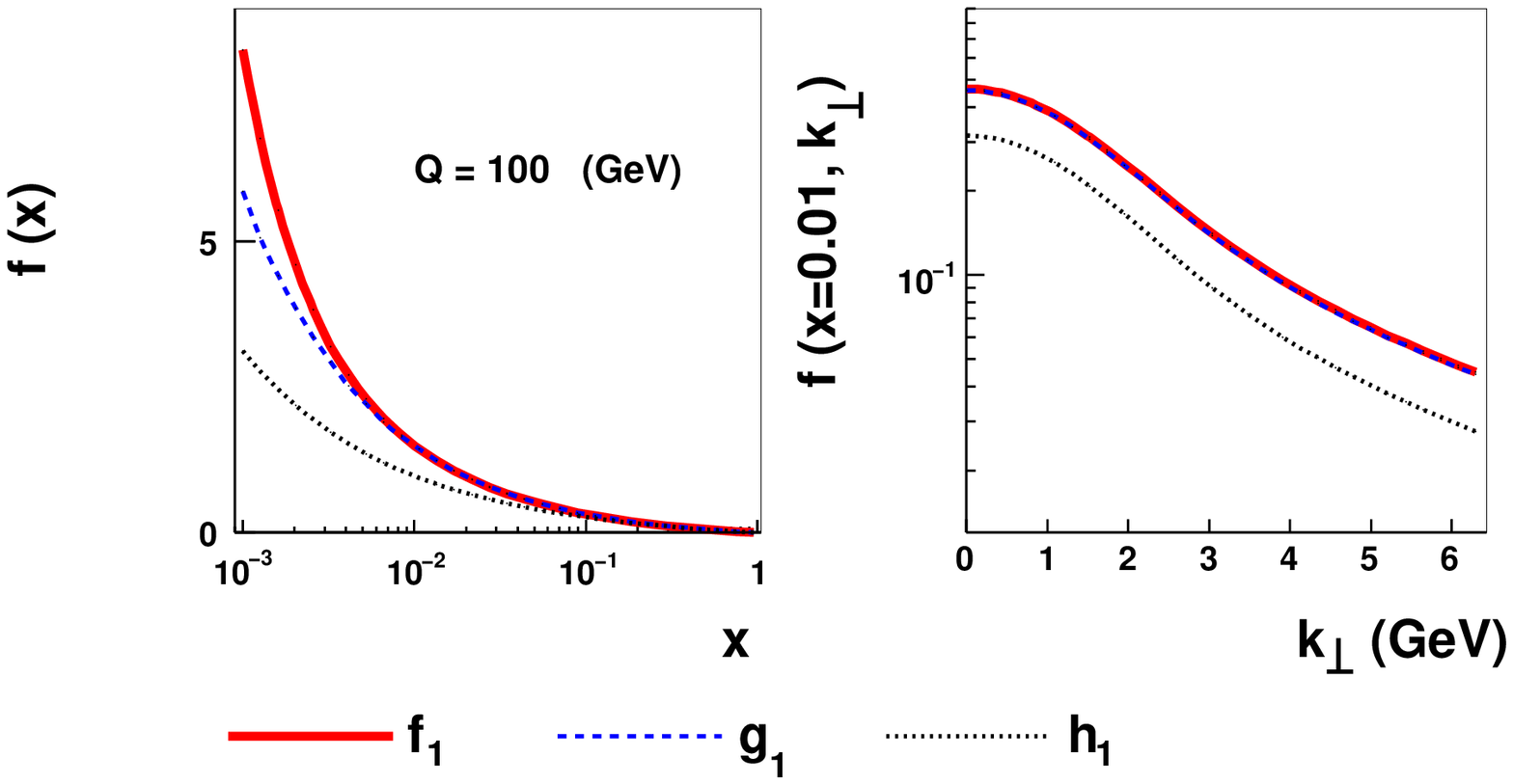, width=0.62\textwidth}
\vskip -2cm
\caption{Evolution of $f^{q}(x,k_T;Q,Q^2)$
and $f^{q}(x;Q,Q^2)$ at two different scales $Q = 3.2$ GeV (two upper plots), $10$ GeV (two middle plots), and $100$ GeV (two bottom plots). Solid line corresponds to unpolarised evolution, dashed line
corresponds to helicity evolution, and dotted line corresponds to transversity evolution.}
\label{fig:comparison3_2}
\end{figure*}
%%%%%%%%%%%%%%%%%%%%%%%%

Let us finally present results of TMD evolution using a test function at
initial scale $Q_0 = 3.2$ GeV. We choose the collinear functions 
in the following form:
\begin{align}
 x u_0(x) = x d_0(x) &\equiv x^{0.5} (1-x)^{0.5} , 
&
 x \bar{u}_0(x) = x \bar{d}_0(x) &\equiv 0 ,\label{eq:initial}
&
 x g_0(x) &\equiv x^{0.5} (1-x)^{0.5} .
\end{align}
Note that the choice is arbitrary as our goal is just to demonstrate the 
results of evolution. 
These initial distributions will be assumed the same for unpolarised ($f_1$), helicity ($g_1$), and transversity ($h_1$) distributions.
Results with realistic 
initial functions will be presented elsewhere.

Evolution in $b_T$ is identical for all three functions, 
see Eq.~\eqref{e:TMDevol1}. 
The convolution in Eq.~\eqref{e:TMDevol1} takes three different forms.
For the unpolarised distribution, 
using Eqs.~(\ref{e:Cffinal}, \ref{e:Cgffinal}) we find
\begin{align}
\sum_j\bigl( 
C_{j/i}\otimes
f^j_1\bigr)(x,b_{\ast};\mu_b)  = f_1(x,\mu_b) +  \frac{\alpha_s C_{\rm F}}{2 \pi} 
\;\int_x^1\frac{d\hat{x}}{\hat{x}} \left(  1 -\frac{x}{\hat x} \right) f_1(\hat{x},\mu_b)+  \frac{\alpha_s T_{f}}{\pi} 
\;\int_x^1\frac{d\hat{x}}{\hat{x}} \frac{x}{\hat x} \left(  1 -\frac{x}{\hat x} \right) g(\hat{x},\mu_b)\;,
\label{eq:evolvf}
\end{align}
where $ g(\hat{x},\mu_b)$ corresponds to the gluon distribution.

For the helicity distribution, using Eqs.~(\ref{e:Cgfinal}, \ref{e:Cggfinal})
we obtain
\begin{align}
\sum_j\bigl( 
\Delta C_{j/i}\otimes
g_1^j\bigr)(x,b_{\ast};\mu_b)  = g_1(x,\mu_b) +  \frac{\alpha_s C_{\rm F}}{2 \pi} 
\;\int_x^1\frac{d\hat{x}}{\hat{x}} \left(  1 -\frac{x}{\hat x} \right) g_1(\hat{x},\mu_b)+  \frac{\alpha_s T_{f}}{\pi} 
\;\int_x^1\frac{d\hat{x}}{\hat{x}} \left(  1 -\frac{x}{\hat x} \right) \Delta g(\hat{x},\mu_b)\; .
\label{eq:evolvg}
\end{align}

Finally, for the transversity distribution, 
using Eqs.~(\ref{e:Chfinal}, \ref{e:Cghfinal}) we obtain
\begin{align}
\sum_j\bigl( 
\delta C_{j/i}\otimes
h_1^j\bigr)(x,b_{\ast};\mu_b)  = h_1(x,\mu_b)\; .
\label{eq:evolvh}
\end{align}

Note that, as is well known for CSS resummation, at this order TMD evolution
does not have mixing of different flavors and no mixing with gluon TMD, even though
collinear PDFs mix as can be seen from Eqs.~(\ref{eq:evolvf},\ref{eq:evolvg},\ref{eq:evolvh}). This
means also that the gluon TMD cannot be studied via
scaling violations at least at this order. 
 
We perform DGLAP evolution for the 
collinear functions $f_1(x,\mu_b), g_1(x,\mu_b), h_1(x,\mu_b)$ using the 
HOPPET evolution package~\cite{Salam:2008qg}.

The choices of non-perturbative functions that enter in Eq.~\eqref{e:TMDevol1}
are the following (we use again the same function for all three polarization
cases just for illustration purposes):
\begin{align} 
 \hat{f}_{ {\rm NP}}^q(x,b_T) &= \exp\left(-\frac{b_T^2 \langle k_T^2
   \rangle}{4 }\right)  \; ,
&
g_K(b_T) &= - g\, b_T^2 \; ,
\label{eq:nonpert}
\end{align}
where $\langle k_T^2 \rangle = 0.25$ (GeV$^2$), $g=0.2$ (GeV$^2$). We also
choose $b_{max} = 1.5$ (GeV$^{-1}$). 

The choice of $\hat{f}_{ {\rm NP}}^q$ corresponds to non-perturbative
functions used in analysis of TMD functions at tree-level 
by the Torino-Cagliari-JLab group~\cite{Anselmino:2007fs,Anselmino:2008sga}. 
The choice of  $g_K(b_T)$ is motivated by the
so-called LBNY fit
of Drell-Yan cross-sections using CSS
resummation formalism~\cite{Landry:2002ix}.

We will show the evolution of the TMD functions, as well as their
integral
over $k_T$ up to the value of $Q$, which we conventionally referred to as
their $0^{\rm th}$ $k_T$-moment: 
\begin{align}   
f^q(x,k_T;\mu,\zeta_F) \;,
&
&
f^q(x;\mu,\zeta_F) &\equiv 2\pi \int_0^{\mu} k_T d k_T f_{q/P}(x,k_T;\mu,\zeta_F) \; .
\label{eq:0moment}
\end{align}
Note that this $0^{\rm th}$ $k_T$-moment af a TMD functions $f^q(x;\mu,\zeta_F)$ should not be confused with collinear
PDF $f^q(x;\mu)$.

In Fig.~\eqref{fig:comparison3_2} we show results of the 
evolution of $f^q(x,k_T;Q,Q^2)$
and $f^q(x;Q,Q^2)$ at three different scales $Q = 3.2$, $10$ and $100$ GeV,
for the unpolarized, helicity and transversity distributions. 
As one can see from Fig.~\eqref{fig:comparison3_2}, after evolution the three
functions become wider and are still very similar.
The only
appreciable differences are at higher transvese momentum.
The $0^{\rm th}$ $k_T$-moments after evolution show some differences at low
$x$. 
Compare our results with results of collinear evolution of $g_1$ and $h_1$ presented, for example in Ref.~\cite{Barone:2001sp}. As in the case of collinear evolution, also in TMD evolution $h_1$ becomes smaller than $g_1$ under evolution and the difference grows with $Q$. The reason is the absence of $\alpha_s^1$ contributions to coefficient functions of transversity.   

We also note that TMD functions at initial scale are very much similar to TMD functions parametrized at tree level \cite{Anselmino:2007fs,Anselmino:2008sga} which thus justify extraction of those functions at tree level. Observables at different 
characteristic scales however should be described using TMD evolution.

 One can also observe that the so called Soffer bound \cite{Soffer:1995ww}:
%\begin{align}
$|h_1(x,Q^2)| \le \frac{1}{2}\left(f_1(x,Q^2) + g_1(x,Q^2) \right)\,$ ,
%\end{align}
is also satisfied for TMD distributions $f^{q}(x,k_T;Q,Q^2)$
and $f^{q}(x;Q,Q^2)$ numerically. Let us remind that Soffer bound for
collinear densities  was shown to be preserved at LO accuracy in
Ref.~\cite{Barone:1997fh} and at NLO accuracy in
Ref.~\cite{Vogelsang:1997ak}. We set aside the discussion of Soffer bound for
TMD functions for a separate publication.

%%%%%%%%%%%%%%%%%%%%%%%%%%%%%%%%%%%%%%%%%%%%%%%%%%%%%%%%%%%%%%%%%%
\section{Conclusions}

In this paper we calculated the evolution of the transverse-momentum-dependent
(TMD) 
helicity and transversity distribution functions. We
adopted the definition of TMD PDFs as given 
by Collins in Ref.~\cite{Collins:2011zzd}. 
We provided explicit formulas for all coefficient functions at $\alpha_S$.
The results of this paper can be readily
used in TMD phenomenology. 

As an illustration, we calculated the unpolarized, helicity and transversity 
TMD distributions at different scales,
starting from the same initial conditions. The final results are very similar.
Their $0^{\rm th}$ $k_T$-moments differ at low $x$.
We observed that if started from
equal initial conditions, helicity TMD distribution $g_1$ becomes smaller than
unpolarised $f_1$ distribution and transversity $h_1$ becomes smaller than
helicity $g_1$ TMD.

%%%%%%%%%%%%%%%%%%%%%%%%%%%%%%%%%%%%%%%%%%%%%%%%%%%%%%%%%%%%%%%%%%

\begin{acknowledgments}
We would like to thank Ted Rogers for multiple discussions, help and encouragement during  
writing of this paper. Authored by a Jefferson Science Associate, LLC under U.S. DOE Contract 
No. DE-AC05-06OR23177.  A.~Bacchetta is partially supported by the Italian MIUR through the PRIN
2008EKLACK, and by the European Community through the Research Infrastructure
Integrating Activity ``HadronPhysics2" (Grant Agreement n. 227431)  
under the $7^{\rm th}$ Framework Programme.    
\end{acknowledgments}

%%%%%%%%%%%%%%%%%%%%%%%%%%%%%%%%%%%%%%%%%%%

%\bibliographystyle{myrevtex}
%\bibliography{mybiblio}

\end{document}